\documentclass[twocolumn]{aastex631}

\usepackage{amsmath}
\usepackage{multirow}
\usepackage{enumitem}
\usepackage{xspace}  

\def\hda{\ensuremath{\rm{H\delta_A}}\xspace}
\def\DF{\ensuremath{\rm{D_{N}4000}}\xspace}

\def\B{\ensuremath{B_{435}}\xspace}
\def\V{\ensuremath{V_{606}}\xspace}
\def\I{\ensuremath{I_{814}}\xspace}
\def\Y{\ensuremath{Y_{105}}\xspace}
\def\J{\ensuremath{J_{125}}\xspace}

\def\H{\ensuremath{H_{160}}\xspace}

\newcommand{\prospector}{{\sc Prospector}\xspace}

\def\sigstar{\ensuremath{\sigma^\prime_{\rm{star}}}\xspace}
\def\q{\ensuremath{q}\xspace}

\begin{document}

\title{Reconstructing the Assembly of Massive Galaxies. III:\\
Quiescent Galaxies Loose Angular Momentum as They Evolve in a Mass-dependent Fashion.
}
\correspondingauthor{Zhiyuan Ji}
\email{zhiyuanji@arizona.edu}

\author[0000-0001-7673-2257]{Zhiyuan Ji}
\affiliation{Steward Observatory, University of Arizona, 933 N. Cherry Avenue, Tucson, AZ 85721, USA}

\author[0000-0002-7831-8751]{Mauro Giavalisco}
\affiliation{University of Massachusetts Amherst, 710 North Pleasant Street, Amherst, MA 01003-9305, USA}

\begin{abstract}

We study the evolution of stellar kinematics of a sample of 952 massive quiescent galaxies with $M_*>10^{10.5}M_\sun$ at $0.6<z<1$. Utilizing spatially integrated spectroscopy from the LEGA-C survey, we focus on the relationship between the observed integrated stellar velocity dispersion (\sigstar) and the morphological axial ratio (\q), and its variation with the stellar age and mass of quiescent galaxies. For the youngest quiescent galaxies, regardless of stellar mass, \sigstar decreases with increasing \q, a trend that is consistent with a system having significant rotation and hence suggests that massive galaxies still retain significant amount of angular momentum in the aftermath of quenching. As they continue to evolve, the variation of the \sigstar-\q relationship depends on stellar mass. For quiescent galaxies with $M_*<10^{11.3}M_\sun$, \sigstar decreases with \q in all stellar-age bins, suggesting that the quiescent populations of this mass regime retain significant rotation even long time after quenching. In contrast, for more massive quiescent galaxies with $M_*>10^{11.3}M_\sun$, the relationship between \sigstar and \q becomes significantly flattened with increasing stellar age. This indicates that, as the very massive galaxy populations continue to evolve after quenching, angular momentum gradually reduces, which eventually transforms them into velocity-dispersion supported systems. We suggest that incoherent, continuous merging and accretion events onto the galaxies are the main drivers of the observed mass-dependent, posting-quenching dynamical evolution, because more massive galaxies are more likely to undergo such interactions. We are witnessing the early formation epoch of fast and slow rotators at $z \sim 0.8$, when the Universe was only half of its age nowadays.

\end{abstract}

\keywords{Galaxy formation(595); Galaxy evolution(594); Galaxy structure(622); High-redshift galaxies(734)}

\section{Introduction} \label{sec:intro}

Early type/quiescent galaxies that no longer actively form stars dominate the cosmic stellar-mass budget in the present-day Universe \citep{Muzzin2013}.  At redshift $z\sim0$, integral-field spectroscopic (IFS) observations revealed a bimodal distribution in the stellar kinematics of massive quiescent galaxies \citep{Emsellem2004, Emsellem2007, Cappellari2007}. Two classes -- fast and slow rotators -- are identified to have distinct  $V/\sigma$, i.e. the ratio of the ordered ($V$) to random ($\sigma$) motions in a stellar system.  Relative to fast rotators, slow rotators have lower $V/\sigma$, and they generally are more massive with stellar masses of $M_*\gtrsim 10^{11.3}M_\odot$ and weakly triaxial \citep{Cappellari2016}. Constraining the pathway to establishing the observed kinematical dichotomy at $z\sim0$ is the key for understanding the assembly of massive galaxies, which requires us to push the study of stellar kinematics in quiescent galaxies towards higher redshifts, i.e. closer to the epoch when the dichotomy was emerging.

For quiescent galaxies  at high redshifts, however, measuring stellar kinematics with spatially resolved spectroscopy is very challenging, owing to their compact morphologies \citep[e.g.][]{vanderWel2014, Ji2024}, and the lack of strong emission lines. Prior to the launch of James Webb Space Telescope (JWST, \citealt{Gardner2023}), such measurements only exist for a handful of rare, extremely bright quiescent galaxies whose observed fluxes are highly magnified due to strong gravitational lensing \citep{Newman2015, Toft2017, Newman2018b}. The immense gain of JWST in sensitivity and angular resolution at IR wavelengths now enables spatially resolved spectroscopy of more general populations (e.g., unlensed) of high-z quiescent galaxies \citep{DEugenio2023}. However, such observations are still time-consuming, requiring $\gtrsim 10$ hours on-source exposure with NIRSpec/IFS \citep{Jakobsen2022} for a single fairly bright ($K<22.5$ mag, $M_*\sim10^{11}M_\sun$) quiescent galaxy \citep{Nanayakkara2022}. This makes it not possible -- even with JWST -- to measure stellar kinematics with spatially resolved spectroscopy in statistically large samples of high-z quiescent galaxies on a rapid timescale.

Yet, notwithstanding the very limited sample size of high-z quiescent galaxies with robust measures of stellar kinematics, the findings from existing studies are somewhat surprising. All systems that have been studied show rapid rotation \citep{Newman2015, Toft2017, Newman2018b,DEugenio2023}, despite that their large stellar mass, i.e. typically $\gtrsim10^{11.3}M_\odot$, suggests that they should be the progenitors of $z\sim0$ slow rotators. If those systems are good representative of the underlying population of high-z massive quiescent galaxies, the implication will be profound: Significant dynamical transformations, particularly the loss of angular momentum, must happen after the quenching of massive galaxies. Unfortunately, such an implication can be fraught with systematic errors, considering that the current sample size is rather small and the sample selection function can be complex for observations on target basis. 

\begin{figure*}
    \centering
    \includegraphics[width=0.7\textwidth]{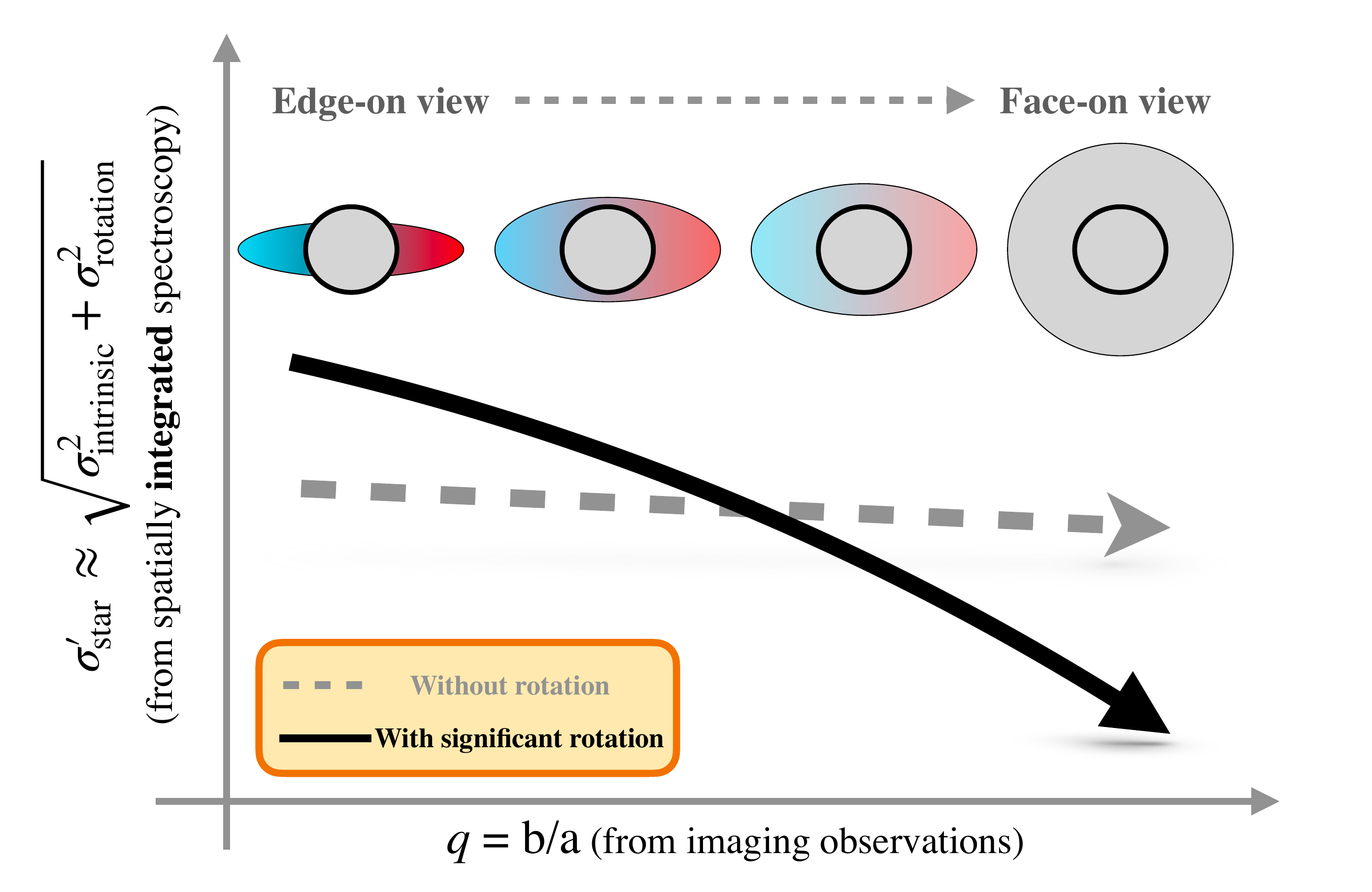}
    \caption{Illustration of the main idea of this work (see Section \ref{sec:intro} for details). We investigate the dynamical transformation of massive quiescent galaxies at high redshifts with spatially integrated/unresolved stellar kinematics by studying the empirical relationship between \sigstar ($y$-axis) and \q ($x$-axis).}
    \label{fig:illustration}
\end{figure*}

In this work, instead of relying on spatially resolved spectroscopy, we investigate the dynamical transformation of quiescent galaxies using spatially integrated/unresolved stellar kinematics. In such a way we are able to conduct the analysis with a statistically significant sample of $\approx1000$ massive quiescent galaxies at $z\sim0.8$, about half the Hubble time of the Universe today. In particular, we focus on the stellar-age dependence of the empirical relationship between \sigstar\footnote{Following the notation of the LEGA-C survey, we use \sigstar to differentiate it from the intrinsic stellar velocity dispersion $\sigma_{\rm{star}}$ commonly used in the literature.}, i.e. the observed integrated stellar velocity dispersion (after taking into account the instrumental resolution), and \q, i.e. the ratio of the semi-minor to semi-major axes of the morphology of galaxies which is a sensitive probe of inclination. The idea is illustrated in Figure \ref{fig:illustration} and described in detail in what follows.

For spatially integrated spectroscopy, both random and ordered (if present) motions contribute to the broadening of intrinsic stellar templates. Therefore, \sigstar equals to the square root of the quadratic sum of the intrinsically random motion ($\sigma$) and the contribution from projected rotation along the line of sight ($\sigma_{\rm{rotation}}$). For a system with significant rotation, because $\sigma_{\rm{rotation}}$ decreases with increasing \q (from edge-on to face-on), \sigstar decreases with increasing \q (black solid line in Figure \ref{fig:illustration}). In contrast, for a system dominated by random motion, because the contribution from $\sigma_{\rm{rotation}}$ to \sigstar is negligible compared to $\sigma$, a much weaker relationship between \sigstar and \q is expected. A very similar idea\footnote{Instead of \sigstar, \citet{Belli2017} studied the relationship between \q and the dynamical mass of galaxies, $M_{\rm{dyn}}$ whose estimation heavily depends on \sigstar measured from the spatially integrated stellar kinematics.} has  been discussed and utilized in an earlier study by  \citet{Belli2017} of a much smaller sample of 24 quiescent galaxies at $z\sim2$. 

With spectral energy distribution (SED) modeling growing in sophistication and accuracy, statistically reconstructing high-fidelity star formation histories (SFHs) is becoming possible for high-z massive galaxies when high-quality, panchromatic data are available. The flexibility of the SFH treatment in SED modeling ensures a much less biased, if at all, inference of physical parameters \citep{Carnall2019, Leja2019}. Built upon this latest development in SED modeling, in the first two papers of this series \citep{paperi, paperii}, we have utilized the fully Bayesian SED fitting code \prospector \citep{Johnson2021} to reconstruct the nonparametric SFH of massive galaxies at $z\sim2$. Combining together the SFHs and morphological analysis, we were able to reconstruct the timing sequence of the morphological transformation of massive galaxies as they evolve from the main sequence to quiescence. 

In this third paper, we focus on the dynamical transformation of quiescent galaxies in approximately the last half of the Hubble time. With robust stellar-age estimates from SED fitting, we study the dependence of the relationship between \sigstar and \q on the stellar age of galaxies. Any significant change of the \sigstar vs. \q relationship with stellar age is an indication of strong evolution in the dynamical state of massive galaxies after they quench. The redshift range considered here is $0.6<z<1$, where statistically significant samples of quiescent galaxies with unresolved stellar kinematics are available. Throughout this paper, we adopt the AB magnitude system and the $\Lambda$CDM cosmology with \citealt{Planck2020} parameters, i.e., $\Omega_m = 0.315$ and $\rm{h = H_0/(100\, km\,s^{-1}\,Mpc^{-1}) = 0.673}$.

\section{The Sample} \label{sec:sample}

The parent sample considered in this study comes from the Large Early Galaxy Astrophysics Census (LEGA-C, \citealt{vanderWel2016,Straatman2018}), the latest and final Data Release 3 \citep{vanderWel2021}. The LEGA-C survey is an ESO/Very Large Telescope public survey that observed with deep spectroscopy (median S/N $\sim15$ at 4000 \AA) for a sample of $\sim3500$ galaxies at $0.6<z<1$, selected using the $K$-band flux from the COSMOS/UltraVISTA survey \citep{Muzzin2013}. Here we only focus on the galaxies with $M_*>10^{10.5}M_\sun$, to ensure (1) good stellar mass completeness (see Figure A1 of \citealt{vanderWel2021}) and (2) that the environmental effects -- external to the host halo of a galaxy -- on the evolution of galaxies are minor \citep[e.g.][]{Ji2018}.

We refine the sample selection using the flags from the LEGA-C data release. We require \texttt{FLAG\_MORPH} $=0$, to ensure that during observations the light through the slit is from a single galaxy with a regular morphology, meaning that mergers and galaxies whose LEGA-C spectra are contaminated by adjacent galaxies are excluded from the sample. We also require \texttt{FLAG\_SPEC} $=0$, to exclude the galaxies with clear AGN presence identified by either IR or X-ray observations. These two constraints together ensure the high-quality spectral measures, and eliminate the cases when the interpretation of the dynamical measures becomes complicated. We cross match the LEGA-C catalog with the photometric catalog of COSMOS2020 \citep{Weaver2022}, and finally select quiescent galaxies using the UVJ criteria of \citet{Muzzin2013}. Our final sample contains 952 UVJ-selected quiescent galaxies. 

\begin{figure*}
    \centering
    \includegraphics[width=0.97\textwidth]{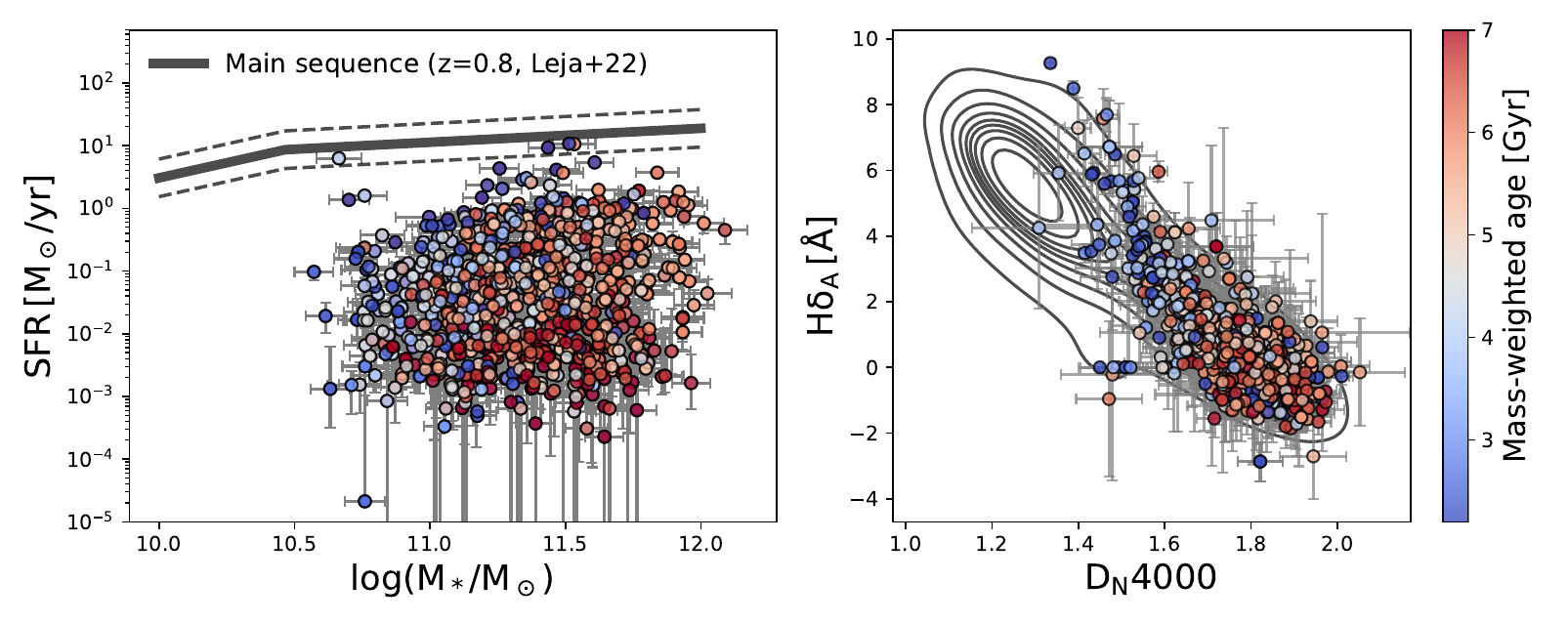}
    \caption{{\bf Left:} Distribution of the final sample of 952 massive quiescent galaxies in the plane of SFR vs $M_*$. The black solid line shows the star-forming main sequence of \citet{Leja2022}, and the black dashed lines mark the range of $\pm0.3$ dex. {\bf Right:} Distribution of the final sample in the plane of \hda vs. \DF. The background grey contours show the distribution of all galaxies with $M_*>10^{10.5}M_\sun$ from the LEGA-C survey. Each one of the quiescent galaxies in our final sample is color coded according to its mass-weighted stellar age derived from SED fitting assuming non-parametric SFH. Our quiescent sample selected via UVJ technique also occupies the region of the parameter space of quiescent galaxies in these two planes.}
    \label{fig:sample}
\end{figure*}

\section{Measurements}

\subsection{\sigstar and \q } 

The measurements of unresolved stellar kinematics \sigstar and morphological axis ratio \q are taken directly from the LEGA-C data release \citep{Bezanson2018b,vanderWel2021}. We refer readers to those references for technical details. 

Briefly, \q is derived from the HST/ACS $I_{814}$ imaging in the COSMOS field, following \citet{vanderWel2012} who used the {\sc Galfit} package \citep{Peng2010} to model the 2D light distribution of galaxies assuming a single S\'{e}rsic profile. \sigstar were measured  with the {\sc pPFX} package \citep{Cappellari2017} by fitting the observed spatially integrated spectra with the combination of (1) high-resolution (R $=$ 10000) theoretical single stellar templates and emission lines at the instrumental resolution, (2) a 3rd-order multiplicative polynomial and (3) an additive polynomial. The unresolved stellar velocity dispersion \sigstar and gas velocity dispersion are estimated independently by broadening the templates with Gaussian kernels. 

\subsection{SED fitting with Prospector} \label{sec:measure}

The properties of the stellar-populations of the sample galaxies are derived by fitting the multi-band  photometry from the COSMOS2020 catalog with the fully Bayesian code \prospector \citep{ Johnson2021}. Each one of the sample galaxies has $\approx40$ band photometry that densely samples the rest-frame UV-to-NIR wavelengths. Compared to previous COSMOS catalogs, COSMOS2020 includes the new, significantly deeper optical and NIR imaging from the Subaru/HSC and VISTA/VIRCAM surveys \citep{Weaver2022}. Two catalogs using different aperture photometric methods are available in the COSMOS2020 release, namely the CLASSIC and FARMER catalogs. By default, we use the former where   aperture-matched photometry was carried out following \citet{Laigle2016}. We note, however, that the difference between the two photometric catalogs is negligible for galaxies in the magnitude range  ($K<21.5$ mag) considered here (Figure 8 and 9 in \citealt{Weaver2022}).

The basic setups of our \prospector fitting are essentially the same as those in the first two papers of this series \citep{paperi, paperii}. We adopt the Flexible Stellar Population Synthesis (FSPS) code \citep{Conroy2009,Conroy2010} where the stellar isochrone libraries MIST \citep{Choi2016,Dotter2016} and the stellar spectral libraries MILES \citep{Falcon-Barroso2011} are used. We assume the \citet{Kroupa2001} initial mass function and the \citet{Byler2017} nebular emission model. We assume the \citealt{Calzetti2000} dust attenuation law and fit the V-band dust optical depth with a uniform prior $\tau_V\in(0,2)$. We fix the redshift to the spectroscopically-measured values from LEGA-C, and set the stellar metallicity as a free parameter with a uniform prior in the logarithmic space $\log(Z_*/Z_\sun) \in (-2, 0.19)$, where the upper limit of the prior is chosen because it is the highest metallicity that the MILES library covers.

We use the nonparametric form of SFH that is critical for unbiased inference of stellar-population properties \citep[e.g.][]{Leja2019}. Specifically, we use a piece-wise step function composed of nine lookback time bins, where the star formation rate (SFR) is constant within each bin. We fix the first two bins as $0-30$ and $30-100$ Myr to capture recent episodes of star formation. We also fix the last bin as 0.9$t_{\rm{H}}-t_{\rm{H}}$ where  $t_{\rm{H}}$ is the Hubble Time of observation. The remaining six bins are evenly spaced in the logarithmic lookback time between 100 Myr and 0.9$t_{\rm{H}}$. 

To ensure the convergence of nonparametric SFH reconstructions and reasonable uncertainty estimations \citep[e.g.][]{Carnall2019,Leja2019}, we adopt the Dirichlet prior \citep{Leja2017} during the \prospector SED fitting. This  prior has been demonstrated to be able to recover the diverse shape of SFHs \citep{Leja2019}. Moreover, using the synthetic observations of simulated galaxies that have similar data quality like the ones we use here for the LEGA-C galaxies, \citet[][see their Appendix A]{paperi} demonstrated that the Dirichlet prior can better recover the stellar age of high-z quiescent galaxies compared to other commonly-used priors, such as the continuity one, which is commonly adopted to measure the SFH of star-forming galaxies. 

In Figure \ref{fig:sample}, we show the distributions of the sample galaxies in the planes of SFR vs. $M_*$, and of  \hda \citep{Worthey1994, Worthey1997} vs. \DF \citep[4000\AA\ break,][]{Balogh1999}. From the left panel of the Figure it is immediately clear that the UVJ-selected quiescent galaxies also occupy the parameter space of galaxies below the star-forming main sequence, i.e. with depressed SFR at any given stellar mass. This shows very good consistency among different selection methods of quiescent galaxies. 

In the right panel of Figure \ref{fig:sample}, each one of the galaxies is color coded according to mass-weighted stellar age.  It has been extensively shown that \hda\ and \DF\ are sensitive diagnostics of galaxy's stellar age \citep[e.g.][]{Kauffmann2003}. As the Figure shows, galaxies with lower SFR (at fixed $M_*$), larger \DF\ and larger (negative) \hda\ also have larger ages (older stellar populations) from our \prospector fitting, demonstrating the robustness of our stellar-age inference. Because the main conclusion of this study depends on the stellar-age measures, in Appendix \ref{app:age} we conduct a number of further tests on the robustness of the age inference. We find that the absolute stellar-age inference can be subject to systematic offsets across different methodologies and calibrations. Such offsets mainly affect the mapping from a given age bin to an absolute timescale, whereas the relative ordering of younger versus older populations at fixed redshift is robust (Appendix \ref{app:age}). Because our analysis relies primarily on relative age ranking, i.e., differential comparisons between age bins, our main conclusions are not sensitive to a uniform shift in the absolute age scale.

\section{Results} \label{sec:result}

We now present the relationship between \sigstar and \q, i.e. the core of this study. We first divide the sample into the low-mass and high-mass subsamples using $M_*=10^{11.3}M_\sun$, i.e. the characteristic mass to separate fast and slow rotators at $z\sim0$ \citep{Cappellari2016}. We then further divide each subsample into three subgroups using the 33th- and 67th- percentiles of the stellar-age distribution of the entire quiescent sample. 

\subsection{The median trend} \label{sec:sigma_q}

We measure the median relationship between \sigstar and \q of each one of the subgroups using the Locally Weighted Scatterplot Smoothing (LOWESS\footnote{www.statsmodels.org/dev/generated/\\statsmodels.nonparametric.smoothers\_lowess.lowess.html}) method that fits a smoothed curve to data points through a non-parametric approach, i.e. the process does not require to assume any specific functional form. We estimate the uncertainty of the median relationship via Monte Carlo simulations. In particular, we use Gaussian distributions to resample the individual  \sigstar and \q measures with their corresponding uncertainties. We then use LOWESS to measure the median \sigstar vs. \q relationship of the resampled data points. We repeat these 1000 times, and use the range between 16th-  and 84th- percentiles as 1-$\sigma$ uncertainty. 

\begin{figure*}
    \centering
    \includegraphics[width=1\textwidth]{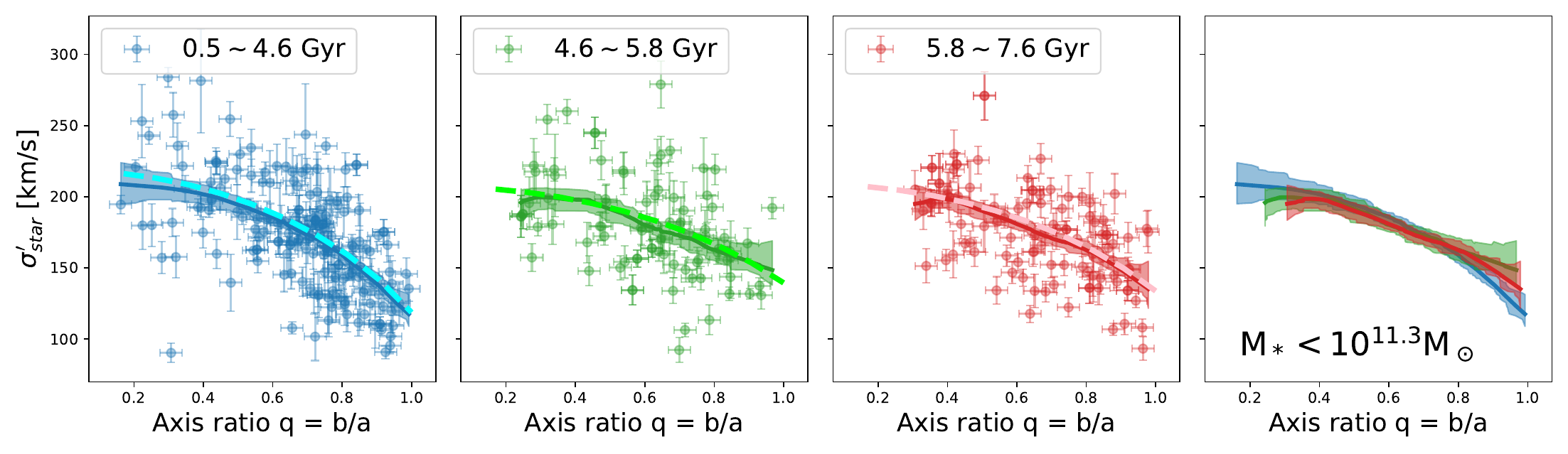}
    \includegraphics[width=1\textwidth]{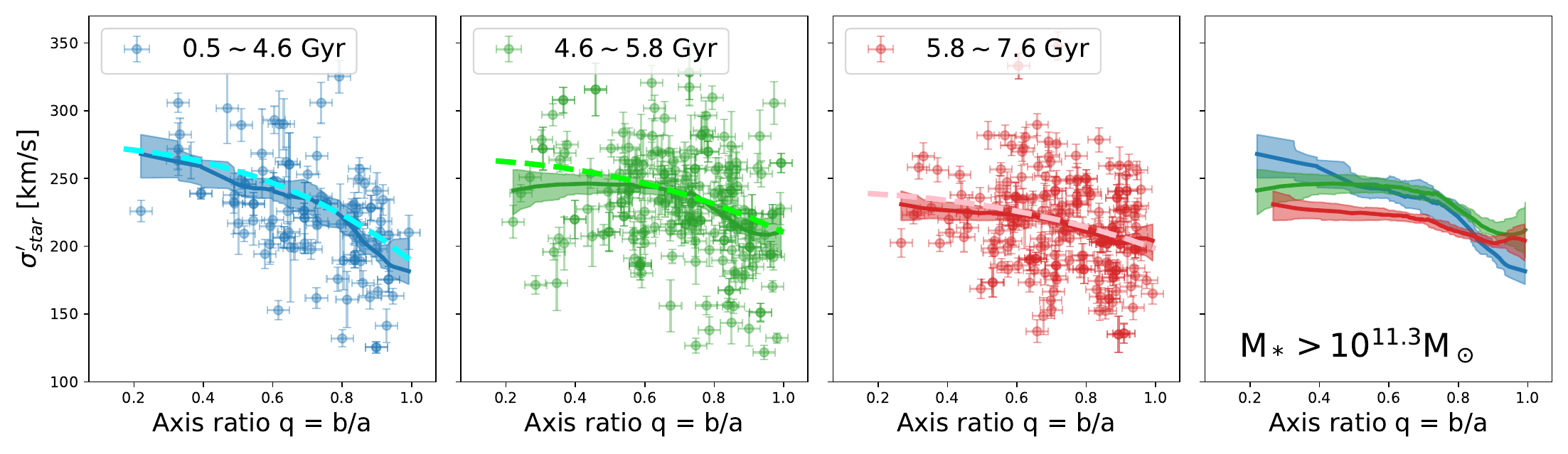}
    \caption{\sigstar vs. \q. The first and second rows show the results of the quiescent galaxies with $M_*<10^{11.3}M_\sun$ and $M_*>10^{11.3}M_\sun$, respectively. For each stellar-mass bin, galaxies are further divided into three age bins, each of which roughly contains 1/3 of the entire sample. In each panel the solid line and shaded region show the median trend and the corresponding 1-$\sigma$ uncertainty derived using the LOWESS algorithm. The dashed line shows the best-fit from our toy model (see Section \ref{sec:model}). In the fourth column, all LOWESS median trends are shown together. }
    \label{fig:sigma_q}
\end{figure*}

To begin, as the first column of Figure \ref{fig:sigma_q} shows, in the youngest age bin, \sigstar decreases with increasing \q, which is observed in both mass bins. This suggests that massive galaxies -- regardless of their masses -- still retain significant rotation in the aftermath of quenching.  As they continue evolving and become older (the second and third columns of Figure \ref{fig:sigma_q}), the relationship between \sigstar and \q starts to differ in the two mass bins. 

For low-mass quiescent galaxies with $M_*<10^{11.3}M_\sun$ (the first row of Figure \ref{fig:sigma_q}), \sigstar decreases with \q in all age bins, suggesting quiescent galaxies in this mass regime continue to retain significant rotation even long time after quenching. Overall, the median relationships between \sigstar and \q are statistically consistent with each other for all age bins within the uncertainties, although there is some evidence that younger galaxies have a steeper relationship than the older ones. 

For high-mass quiescent galaxies with $M_*>10^{11.3}M_\sun$ (the second row of Figure \ref{fig:sigma_q}), the decreasing trend of \sigstar with \q becomes significantly flattened as they become older. This shows that post-quenching dynamical transformations (1) happen in these very massive systems, which significantly reduce the amount of rotation and (2) these transformations are more profound in more massive quiescent galaxies.

In the analysis above we divided the low- and high-mass subsamples into only three age bins. We continue to study the variation of the \sigstar-\q relationship by dividing the subsamples into more age bins. Specifically, instead of binning the subsamples using arbitrary bins of stellar age, we first sort stellar ages of individual galaxies into an increasing order. And starting from the first 30\% of the sorted subsamples, we measure the Spearman's rank correlation coefficient $\rho$. Then, we keep adding older quiescent galaxies into the correlation test, and study the change of $\rho$ as a function of the maximum age of the galaxies included in the measure. The uncertainty of $\rho$ calculated in this way is estimated by bootstrapping the sample galaxies 1000 times, and during each bootstrapping iteration we also resample the values of \sigstar and \q with their corresponding measurement uncertainties using Gaussian distributions.

In Figure \ref{fig:mass_q}, $\rho$ is plotted against the maximum mass-weighted age of the quiescent galaxies included to the Spearman's rank correlation test. For galaxies that are freshly quenched, i.e. with relatively young stellar ages, we find a strong negative ($\rho\sim-0.5$, i.e. a decreasing trend) correlation between \sigstar and \q, regardless of stellar mass. As older quiescent galaxies are added to the correlation test, $\rho$ gradually changes from  $-0.5$ to $-0.3$ for the high-mass ($>10^{11.3}M_\sun$) quiescent populations, while it remains approximately unchanged for the low-mass ones. This shows that the variation of the \sigstar vs. \q relationship with age is significantly stronger in more massive quiescent galaxies, confirming the conclusion reached above based on Figure \ref{fig:sigma_q}. 

\subsection{A toy model for the \sigstar vs. \q relationship} \label{sec:model}

We now introduce a simple toy model, in an attempt to quantify the contribution from rotation to the observed relationship between \sigstar and \q. 

As detailed already in Section \ref{sec:intro}, both random and ordered motions contribute to \sigstar, which can be expressed as
\begin{equation}
	\begin{split}
		(\sigma^\prime_{\rm{star}})^2 = \sigma^2 + \sigma_{\rm{rotation}}^2= \\
		\sigma^2\Big(1 + \gamma^2(\frac{V}{\sigma})^2 sin^2 i\Big).
	\end{split}\label{equ:s}
\end{equation} 
In the above equation we have used $\sigma_{\rm{rotation}} = \gamma V sin\, i$, which describes velocity dispersion observed through a slit (i.e. unresolved spectroscopy) due to a purely rotating disk, where $V$ is the rotational velocity, $i$ is the inclination and $\gamma$ is the conversion factor. Because to our knowledge there is no direct, statistical estimate of $\gamma$ at high redshifts, we decide to fix $\gamma= 0.7$ which is the median value from \citet{Cappellari2013} who measured it using the spatially resolved stellar kinematics of $z\sim0$ early type galaxies from ATLAS$^{\rm{3D}}$. The inclination can be estimated using \q as
\begin{equation}
	sin\, i = \sqrt{\frac{1-q^2}{1-q_z^2}}
\end{equation}
where $q_z$ is the vertical thickness of a disk. Following \citet{Belli2017}, we fix it to be  $q_z=0.2$, which is a reasonable assumption for the following reasons. First, the intrinsic thickness of $z\sim0$ disks inferred from SDSS samples is $\sim0.2$ \citep{Padilla2008}. \citet{vanderWel2014b} extended the analysis of the intrinsic, three-dimensional shape distribution of star-forming galaxies up to $z\sim2.5$, where they found that present-day star-forming galaxies are ``thin, nearly oblate'' systems with intrinsic short-to-long axis ratio $\sim0.25$, and that disks are the most common geometric shape at all $z\lesssim2$ for massive galaxies. Moreover, \citet{HamiltonCampos2023} recently measured rest-optical scale heights for 491 edge-on disks over $0.4<z<2.5$ and found a median intrinsic scale height of $\sim0.7$~kpc with little redshift evolution. While a scale height is not identical to $q_z$, combining this with typical star-forming galaxy sizes at $z\sim0.8$ from \citet{vanderWel2014} implies characteristic thickness-to-size ratios of order $\sim0.1$--$0.3$. Taken together, this argues that adopting $q_z=0.2$ for the toy model at $z\sim0.8$ is reasonable and not an outlier relative to broader constraints on galaxy thickness.

The remaining unknowns in Equation \ref{equ:s} are $\sigma$ and $V$ (or $V/\sigma$) that we attempt to constrain by fitting Equation \ref{equ:s} to the observed \sigstar vs. \q relationship. To estimate the uncertainties of the fitted parameters, we use the same Monte Carlo method mentioned above by resampling the \sigstar and \q measures using their uncertainties. 

The best-fit relationships are plotted as dashed lines in Figure \ref{fig:sigma_q}. The observations can be reproduced very well with the toy model. The best-fit models are in excellent agreement with the LOWESS median trends.

\begin{figure*}
    \centering
    \includegraphics[width=0.7\textwidth]{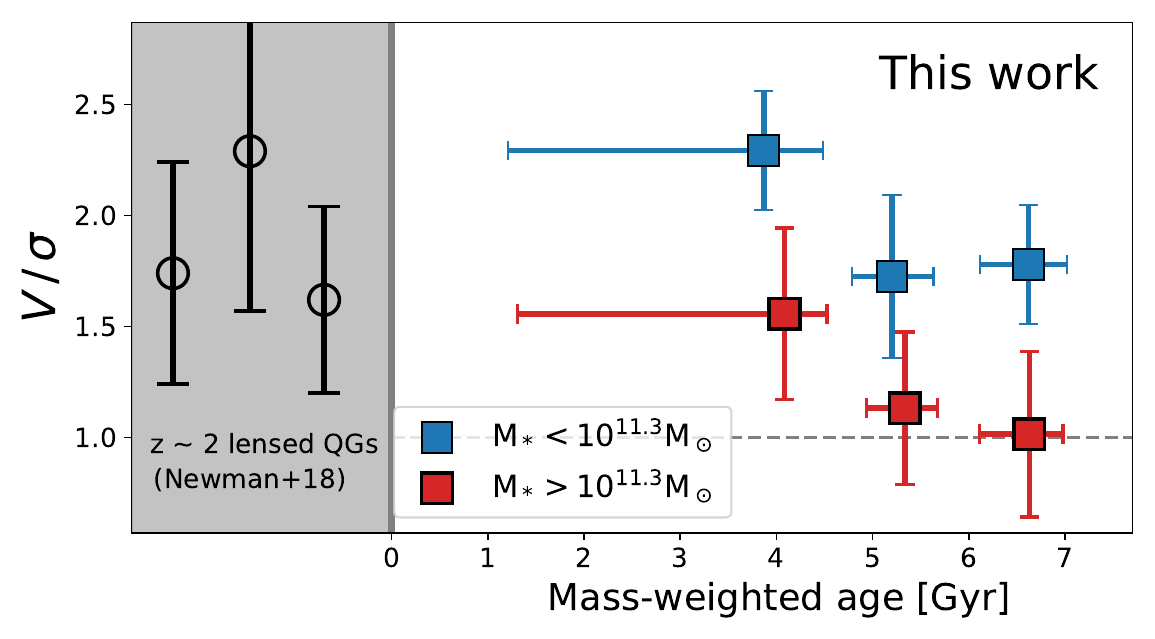}
    \caption{$V/\sigma$ as a function of stellar age. Blue ($M_*<10^{11.3}M_\sun$) and red  ($M_*>10^{11.3}M_\sun$) squares show the $V/\sigma$, inferred from our toy model (Section \ref{sec:model}), of $z\sim0.8$ quiescent populations of different ages. The horizontal dashed line marks $V/\sigma=1$. In the left-most grey shaded region, we also plot three strongly lensed massive ($\gtrsim10^{11.3}M_\sun$) quiescent galaxies at  $z\sim2$ with direct, spatially resolved measures of stellar kinematics from \citet{Newman2018b}.}
    \label{fig:v_2_s}
\end{figure*}

In Figure \ref{fig:v_2_s}, the inferred $V/\sigma$ is plotted as a function of stellar age. Regardless of stellar age, the $V/\sigma$ of the quiescent galaxies with $M_*<10^{11.3}M_\sun$ is greater than 1. As the population evolves, the $V/\sigma$ of these systems decreases from $2.3\pm0.3$ to $1.7\pm0.4$, implying that they remain rotationally supported, with $V/\sigma \sim1.7$, even at least 7 Gyr after their formation\footnote{Note that we do not know the quenching time of the sample galaxies. But, statistically, at a fixed redshift quiescent galaxies formed earlier, i.e. having older stellar ages, should also quench earlier.}. In contrast, during the same cosmic time, the $V/\sigma$ of the higher-mass quiescent galaxies with $M_*>10^{11.3}M_\sun$ monotonically decreases, from $1.6\pm0.3$ (rotationally supported) to $1.0\pm0.3$ (velocity dispersion supported), with increasing stellar age.  

It is important to note several caveats of the $V/\sigma$ inferred from our simple model. The misalignment between the slit and the kinematic major axis of galaxies has not been taken into account in our analysis. Neglecting this effect, however -- given the large sample size of this study -- should lead to an equal/similar systematic error in all subgroups and hence may not cause any substantial impacts on our conclusions, which only rely on a differential comparison. Indeed, we have checked the misalignment angle ($|\mathrm{PA}|$) between the slit and the galaxy semi-major axis. We find mean misalignment angles (with standard deviations) of $44.8\pm26.1^\circ$, $45.4\pm26.5^\circ$, and $48.8\pm26.7^\circ$ for the three stellar-age bins, respectively; these values are consistent with one another within the uncertainties.

Also, the uncertainty in the value of $\gamma$ in Equation~\ref{equ:s} introduces a systematic uncertainty in the inferred $V/\sigma$. Although $\gamma$ cannot be measured directly for our galaxies with the current data, we can assess the likely impact using constraints from the literature and by clarifying which aspects of our results are sensitive to $\gamma$. Adopting the same $\gamma$-related prescription, \citet{Belli2017} inferred slightly higher $V/\sigma$ for $z\sim0$ early-type galaxies from integrated spectroscopy than values obtained from spatially resolved measurements, which may suggest that the effective $\gamma$ for stellar kinematics differs from the fiducial value assumed in simple models. In addition, statistical estimates of $\gamma$ from high-redshift \emph{gas} kinematics typically fall in the range $\sim0.6-1$ (e.g., \citealt{Weiner2006}), but how (and whether) these values translate to \emph{stellar} kinematics remains uncertain. Importantly, the effect of $\gamma$ is primarily to shift the normalization of the inferred $V/\sigma$: an intrinsically larger $\gamma$ would yield a smaller inferred $V/\sigma$. We therefore caution that the absolute $V/\sigma$ values shown in Figure~\ref{fig:v_2_s} are model-dependent. However, unless $\gamma$ varies systematically with stellar age and/or stellar mass across our sample, the qualitative trends reported here, in particular the mass-dependent evolution of $V/\sigma$ with stellar age, should remain robust.

The remaining uncertainty of the inferred $V/\sigma$ from our toy model comes from $q_z$ in Equation~2.  We emphasize that $q_z$ is not directly measured for our sample and may plausibly vary with other physical properties, such that a single fixed value may not apply to all galaxies. Varying $q_z$ within the plausible range above primarily shifts the normalization of the inferred $V/\sigma$ (and contributes additional scatter), and therefore the absolute $V/\sigma$ values should be interpreted with caution. However, unless $q_z$ varies systematically with stellar age (or in a way that correlates with the trends discussed here), the qualitative conclusions of this work, in particular the relative, mass-dependent evolution of $V/\sigma$ with stellar age, should remain robust.

Finally, we also clarify that, when we state, e.g., that the quiescent population of a given stellar age is rotationally supported ($V/\sigma>1$), we do not mean that each one of the quiescent galaxies of that age bin retains significant rotation or that it has a disk, since the spatially integrated spectroscopy does not allow us to constrain that. Instead, what we really mean is that the quiescent {\it population} of that age {\it on average} should have significant rotation.

With the aforementioned caveats in mind, we now compare the inferred $V/\sigma$ of this work with previous studies of the stellar kinematics in quiescent galaxies. \citet{Bezanson2018} pioneered a LEGA-C study of the dynamical transformation of $z\sim0.8$ massive quiescent galaxies using a small (relative to this study) sample of $\sim 100$ galaxies whose major axes are overall aligned with the slit ($|\rm{PA}|<45^{\circ}$), which allows spatially resolved analysis of stellar kinematics. They found that the most massive ($>10^{11.3}M_\sun$) quiescent galaxies show much less rotation compared to less massive systems. In broad agreement\footnote{We do not directly compare our inferred $V/\sigma$ with that of \citet{Bezanson2018}, because with spatially resolved analysis they were able to measure $V_5 /\sigma$ where $V_5$ is the rotation velosity at 5 kpc that we are unable to constrain.} with \citet{Bezanson2018}, our model suggests that quiescent populations of $M_*>10^{11.3}M_\sun$ are significantly less rotationally supported compared to the lower-mass ones, with the inferred $V/\sigma$ for the high-mass subsample being 1.6 times smaller that that for the low-mass one. The $\approx 10\times$ larger in sample size of this study allows us to further group galaxies according to their stellar ages, adding a new piece of information regarding the dynamical transformation of quiescent galaxies as they evolve, as discussed below.


\begin{figure}
    \centering
    \includegraphics[width=0.47\textwidth]{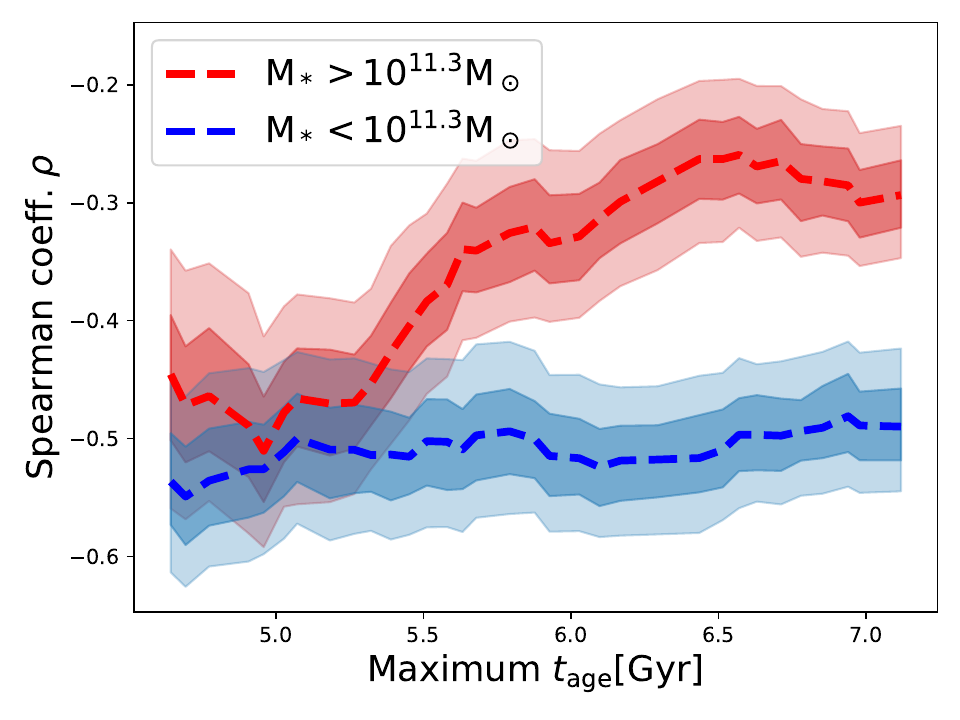}
    \caption{Change of the Spearman's rank correlation coefficient as older quiescent galaxies are included to the correlation test (dashed lines). The $x$-axis shows the maximum mass-weighted age of the quiescent galaxies included in the test. The dark and light shaded regions mark the 0.5- and 1-$\sigma$ uncertainties.}
    \label{fig:mass_q}
\end{figure}

\section{Discussion}
\label{sec:diss}

\subsection{Link to the local fast/slow rotator dichotomy}
Our results can be placed in the broader context of the fast/slow rotator dichotomy established for nearby early-type galaxies using integral-field stellar kinematics (e.g., ATLAS$^{\rm 3D}$; \citealt{Emsellem2011}). While our analysis uses $V/\sigma$ inferred from integrated spectroscopy as a proxy for dynamical support, it is closely related to the local $\lambda_R$-based framework in the sense that both quantify the relative importance of ordered rotation versus random motions. We emphasize, however, that our data do not enable a direct fast/slow-rotator classification (which typically requires spatially resolved kinematics and well-defined apertures), and therefore we are not able to quantify the slow-rotator fraction at $z\sim0.8$.

Recent observational work provides an empirical bridge between $z\sim0.8$ and $z\sim0$. In particular, the MAGPI integral-field survey extends comparable IFS-based classifications to $z\sim0.3$ and, in direct comparisons to local IFS samples, finds that the fraction of slow rotators  is broadly consistent between $z\sim0.3$ and $z\sim0$ \citep{Derkenne2024}. This result suggests that the fast/slow rotator dichotomy, at least for the most massive quiescent galaxies, is already largely established by $z\sim0.3$, leaving relatively limited room for strong additional evolution in the slow-rotator fraction at later times.

Our $z\sim0.8$ analysis is complementary in that it probes earlier stages using population-averaged constraints on rotational support. Although we cannot determine whether the dichotomy is already fully in place at $z\sim0.8$, the clear stellar-mass dependence of the inferred angular-momentum loss in our sample indicates that the physical processes that ultimately produce the local kinematic sequence are already operating by this epoch. In other words, our results are consistent with the dichotomy either (i) already being established by $z\sim0.8$ but not directly measurable with our integrated data, or (ii) still emerging, with the mass-dependent decline in angular momentum and $V/\sigma$ representing an intermediate stage on the path toward the $z\sim0$ fast/slow rotator populations. Distinguishing between these possibilities will require spatially resolved spectroscopy for statistically significant samples at high redshift that can directly measure $\lambda_R$ (or equivalent) and the slow-rotator fraction as a function of mass and stellar age.

\subsection{Connection to spatially resolved kinematics of $z\gtrsim2$ quiescent galaxies}

An important context for our results comes from the small but growing set of spatially resolved stellar-kinematic measurements of quiescent galaxies at $z\sim2$, particularly those enabled by strong gravitational lensing (e.g., \citealt{Toft2017,Newman2018a,Newman2018b}). These studies suggest that at least some very massive quiescent systems at $z\sim2$ remain substantially rotation-supported, with relatively high $V/\sigma$ compared to typical low-redshift quiescent galaxies. In Figure \ref{fig:v_2_s}, we include the measurements from \citet{Newman2018b}, who derived spatially resolved stellar kinematics for three strongly lensed quiescent galaxies with $M_*\gtrsim10^{11.3}\,M_\odot$ at $z\sim2$. With all the systematics in our inferred $V/\sigma$ in mind (Section \ref{sec:model}), on average these $z\sim2$ systems have higher $V/\sigma$ than that inferred for the youngest $z\sim0.8$ quiescent populations of comparable mass.

This comparison provides suggestive constraints on the timing of kinematic transformation. The median stellar age of the youngest bin in our $z\sim0.8$ sample is $\sim3$~Gyr, which is comparable to (and in many cases exceeds) the cosmic age at $z\sim2$. The $z\sim2$ lensed quiescent galaxies must therefore be, on average, more recently quenched than the youngest quiescent populations considered in this study. The fact that such recently quenched, very massive quiescent systems can still retain high rotational velocity implies that quenching does not necessarily coincide with an immediate loss of ordered rotation. Instead, these resolved measurements are consistent with a scenario in which angular-momentum loss is a continuous, cumulative process that can proceed after quenching, progressively reducing net ordered rotation and specific angular momentum as galaxies evolve within the quiescent population.

We therefore interpret our population-level trends in that framework: recently quenched galaxies may enter the quiescent population as fast rotators, while subsequent dynamical evolution, such as cumulative, largely incoherent accretion/merging events and associated heating, gradually reduces $V/\sigma$ and angular momentum. Our finding that angular-momentum loss is more pronounced at higher stellar mass is naturally compatible with these post-quenching processes having a stronger cumulative impact in higher-mass systems. While the current resolved samples are too small to establish a representative evolutionary track at fixed mass, taken together they support the qualitative picture that rotational support in very massive quiescent galaxies can decline over several Gyr after quenching.

Finally, we stress that direct one-to-one comparisons between our measurements and the lensed/resolved samples require care, given the small sample sizes, as well as differences in kinematic estimators and aperture definitions. Nevertheless, the resolved high-$z$ results and the population trends presented here together point toward angular-momentum loss that can continue within the quiescent population and accumulate over time, with a stronger net effect at higher mass.

\subsection{Interpreting the stellar-mass dependence of angular momentum loss in the context of mergers}

Our findings that the post-quenching loss of angular momentum depends on stellar mass are consistent with a scenario in which the most massive quiescent galaxies form in dense environments, likely corresponding to large overdensities in the primordial density field. After quenching, these systems can continue to evolve through gas-poor (dry) mass assembly, either via residual, low-level accretion or through repeated, incoherent mergers with nearby galaxies driven by dynamical friction. Such merger episodes are expected to reduce the specific angular momentum of the remnant and can help produce the observed population of slowly rotating massive quiescent galaxies \citep[e.g.][]{Emsellem2011}.  In contrast, lower-mass quiescent galaxies likely form in regions with comparatively smaller overdensities and reside in shallower potential wells, where the rate of dry merging is substantially lower than in massive halos. As a result, these galaxies experience fewer merger-driven perturbations and can retain significant rotation for long periods after quenching.

Motivated by this picture, we develop a simple toy model to estimate the magnitude of the specific-angular-momentum loss expected from post-quenching dry mergers, detailed below.

\subsubsection{Quantitatively estimating the loss of angular momentum due to dry mergers}
\label{sec:loss_J}
Let the differential merger rate per descendant galaxy with stellar mass $M_*$ be
\begin{equation}
    \lambda(M_*,\mu,z) = \frac{d N_{merg}}{d\mu\, dt}(M_*,\mu,z),
\end{equation}
which is essentially the number of mergers per galaxy per unit time ($t$) per unit mass ratio ($\mu$, defined as the stellar-mass ratio of the secondary to the primary, hence $\mu\le 1$). Consider a short time interval $dt$; the expected number of mergers with mass ratio in $[\mu,\mu+d\mu]$ is then
\begin{equation}
    d\mathbb{E}[N_{merg}] = \lambda(M_*,\mu,z)~d\mu\, dt.
\end{equation}
For each such merger, we assume the galaxy’s specific angular momentum ($j=J/M_*$) changes by a multiplicative factor $F(\mu)$, i.e., $j_{post}=j_{begin}F(\mu)$. Then
\begin{equation}
    \Delta(\ln j) = \ln F(\mu).
\end{equation}
The expected change in $\langle\ln j\rangle$ per unit time is then
\begin{equation}
    \frac{d\,\langle\ln j\rangle}{dt} = \int_{\mu_{min}}^1 d\mu~ \lambda(M_*,\mu,z)\ln F(\mu).
\end{equation}

To derive $F(\mu)$, let's consider two galaxies with angular momentum vectors $\vec{J_1}$ (primary) and $\vec{J_2}$ (secondary). Observations find $j\propto M_*^{2/3}$ \citep{Fall2013}, hence $J\propto M_*^{5/3}$ and
\begin{equation}
    \frac{J_2}{J_1} = \left(\frac{M_2}{M_1}\right)^{5/3} = \mu^{5/3} \equiv q.
\end{equation}
Let $\theta$ be the angle between $\vec{J_1}$ and $\vec{J_2}$. The post-merger vector-sum magnitude is then
\begin{equation}
    |\vec{J}| \equiv |\vec{J_1}+\vec{J_2}| = \sqrt{J_1^2+J_2^2+2J_1J_2\cos\theta},
\end{equation}
Therefore, the expectation over $\theta$ is
\begin{equation}
    \langle|\vec{J}|\rangle  = \int_{0}^{\pi} \sqrt{J_1^2+J_2^2+2J_1J_2\cos\theta}~p(\theta)\,d\theta.
\end{equation}
Assuming the orientation between the two galaxies is random (i.e., isotropic), the probability of landing in a thin ring on a sphere between $\theta$ and $\theta+d\theta$ is proportional to the ring area. One can show\footnote{$p(\theta)d\theta = \frac{2\pi \sin\theta\, d\theta}{4\pi}=\frac{1}{2}\sin\theta\, d\theta$}
\begin{equation}
    p(\theta) = \frac{1}{2}\sin\theta.
\end{equation}
Plugging $p(\theta)$ back into Equation (9) and letting $x=\cos\theta$, we get
\begin{equation}
    \langle|\vec{J}|\rangle = \frac{J_1}{2}\int_{-1}^{1} \sqrt{1+q^2+2qx}~dx.
\end{equation}
Substituting $u=1+q^2+2qx$, we get
\begin{equation}
    \langle|\vec{J}|\rangle = \frac{J_1}{4q}\int_{(1-q)^2}^{(1+q)^2}u^{1/2}\, du,
\end{equation}
and
\begin{equation}
    \langle|\vec{J}|\rangle = J_1\left(1+\frac{q^2}{3}\right) = J_1\left(1+\frac{\mu^{10/3}}{3}\right).
\end{equation}
Finally, since the total stellar mass after the merger is $M_1+M_2 = (1+\mu)M_1$, we get the specific angular momentum after the merger to be
\begin{equation}
    j_{post} = \frac{\langle|\vec{J}|\rangle}{(1+\mu)M_1} = \frac{J_1}{M_1}~\frac{1+\frac{\mu^{10/3}}{3}}{1+\mu},
\end{equation}
and hence
\begin{equation}
    F(\mu) = \frac{1+\frac{\mu^{10/3}}{3}}{1+\mu}.
\end{equation}
Since $0<\mu\le1$, it is obvious that $\ln F(\mu)<0$.

Now, the only remaining unknown in Equation~(6) is $\lambda(M_*,\mu,z)$, which we take from cosmological hydrodynamical simulations. We note that the dominant driver of merger statistics is collisionless gravity. Large $N$-body analyses show that halo merger rates follow near-universal scalings and can be fit to $\sim$10-20\% accuracy over broad ranges of mass, mass ratio, and redshift \citep[e.g.,][]{Fakhouri2010}. Here, we adopt the calibration from \citet[][their Table~1]{RodriguezGomez2015}, which is widely used in the literature and is in good agreement with observational constraints \citep[e.g.,][]{Conselice2022}. It has the functional form
\begin{equation}
\begin{split}
\frac{\lambda(M_*,\mu,z)}{\rm{[Gyr^{-1}]}} &= A(z)\left(\frac{M_*}{10^{10}\,M_\odot}\right)^{\alpha(z)} \left[1+\left(\frac{M_*}{M_0}\right)^{\delta(z)}\right]\,\\
& \mu^{\,\beta(z)+\gamma\log_{10}\!\left(\frac{M_*}{10^{10}\,M_\odot}\right)}. 
\end{split}
\end{equation}
where
\begin{equation}
\begin{split}
A(z) = A_0(1+z)^{\eta}~\rm{with}~(A_0,\eta)=(10^{-2.23},2.46), \\
\alpha(z) = \alpha_0(1+z)^{\alpha_1} ~\rm{with}~(\alpha_0,\alpha_1)=(0.22,-1.18), \\
\beta(z) = \beta_0(1+z)^{\beta_1}~\rm{with}~(\beta_0,\beta_1)=(-1.26,0.06), \\
\delta(z) = \delta_0(1+z)^{\delta_1}~\rm{with}~(\delta_0,\delta_1)=(0.77,-0.47), \\
\gamma = -0.05~\rm{and}~M_0 = 2\times10^{11}M_\odot.
\end{split}
\end{equation}

With $F(\mu)$ and $\lambda(M_*,\mu,z)$, we can finally calculate the specific angular-momentum loss rate as a function of stellar mass, i.e., using Equation 6, where $\mu_{\rm min}$ is fixed to be $1/1000$, the lower limit down to which the fitting formula from \citet{RodriguezGomez2015} is calibrated. We verified that our conclusions would not substantially changed even if $\mu_{\rm min}=1/10$ is used.

\subsubsection{Instantaneous loss rate of specific angular momentum}
\label{sec:J_loss_rate}
\begin{figure*}
    \includegraphics[width=1\textwidth]{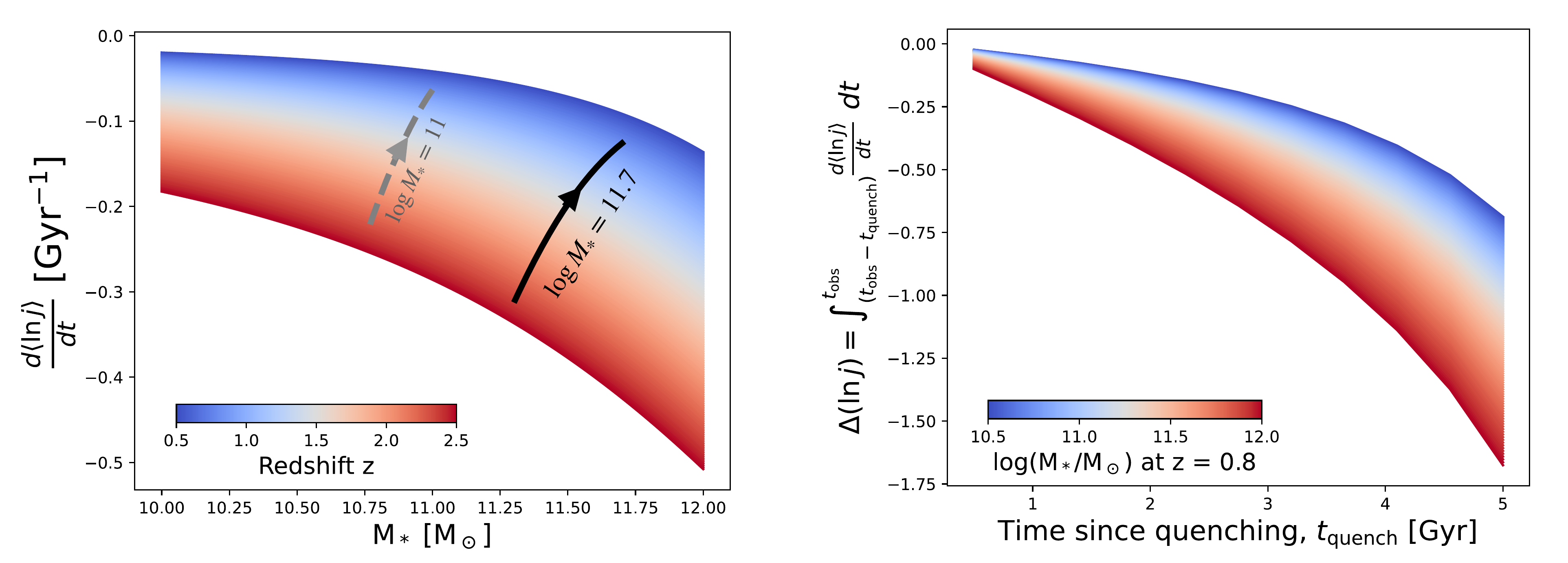}
    \caption{Loss of specific stellar angular momentum ($j$) in our toy model of post-quenching, random-orientation dry mergers (see Section~\ref{sec:loss_J}). \textbf{Left:} Instantaneous loss rate at different redshifts, $d\langle \ln j\rangle/dt$, as a function of descendant stellar mass $M_*$. The black and gray arrows mark representative tracks of merger-driven mass growth for $\log M_*|_{z=0.8}=11.7$ and $11.0$, corresponding to the median values of the two mass bins shown in Figure~\ref{fig:sigma_q}. \textbf{Right:} Cumulative angular-momentum change since quenching, $\Delta(\ln j)$, shown as a function of time since quenching $t_{\rm quench}$, with colors denoting stellar mass at $z=0.8$. More massive systems exhibit faster declines in $j$ and therefore larger (more negative) $\Delta(\ln j)$ over a fixed quenching timescale.}

    \label{fig:J_model}
\end{figure*}

The left panel of Figure~\ref{fig:J_model} shows the predicted instantaneous loss rate of specific angular momentum,
$d\langle \ln j\rangle/dt$ (Equation 6), in our toy model. At fixed redshift, the loss rate becomes
increasingly negative with increasing descendant stellar mass, implying that more massive quenched galaxies are expected
to lose their specific angular momentum more rapidly through subsequent dry mergers. This trend is a direct consequence
of the mass dependence of the merger rate $\lambda(M_*,\mu,z)$ (Equation 16) that enters the integral in Equation 6. In particular,
Equation 16 makes explicit that the dominant mass scaling comes from 
$M_*^{\alpha(z)}\,[1+(M_*/M_0)^{\delta(z)}]$, whereas the residual mass dependence in the $\mu$-term is weak
because $\gamma\ll 1$. Since $\ln F(\mu)<0$, a higher merger rate translates directly into a larger-magnitude (more negative)
$d\langle \ln j\rangle/dt$.

This scaling can be easily understood in two limiting regimes. When $M_*\ll M_0$, the bracketed factor in Equation~16 is
approximately unity and the merger rate scales as $\lambda\propto M_*^{\alpha(z)}$. In this regime, the predicted loss
rate inherits the same (relatively shallow) power-law dependence on stellar mass, and thus the increase in the magnitude of the
loss rate, $|d\langle \ln j\rangle/dt|$, with $M_*$ is modest. In contrast, when $M_*\gg M_0$, the term
$[1+(M_*/M_0)^{\delta(z)}]$ asymptotes to $(M_*/M_0)^{\delta(z)}$, yielding $\lambda\propto M_*^{\alpha(z)+\delta(z)}$.
The corresponding $d\langle \ln j\rangle/dt$ therefore steepens substantially at the high-mass end, producing the
pronounced separation between the low- and high-mass tracks seen in the left panel. We note that the transition occurs near $M_0= 2\times 10^{11}\,M_\odot$, which is similar to the stellar-mass
division adopted for our sample ($ 10^{11.3}\,M_\odot$). This is also the scale at which we observe a clear
stellar-mass dependence of the angular-momentum loss in Figure \ref{fig:sigma_q}. In the context of Equation 16, this
corresponds to the regime where the additional $(M_*/M_0)^{\delta(z)}$ term begins to dominate the merger-rate
normalization. Physically, this implies that the most massive quenched galaxies undergo a higher overall frequency of mergers and,
consequently, experience more frequent merger events that are effective at reducing $j$ under random orientations. The
net result is an accelerated decline of specific angular momentum with time compared to lower-mass systems.

\subsubsection{Cumulative angular-momentum change after quenching}
We now use our toy model to estimate the cumulative change in specific angular momentum after quenching:
\begin{equation}
    \Delta (\ln j) = \int_{(t_{\rm{obs}} - t_{\rm{quench}})}^{t_{\rm{obs}}} \frac{d\langle \ln j\rangle}{dt}~dt.
\end{equation}
where $t_{\rm quench}$ is the time since quenching, defined as the time interval between the quenching time and the observation time. Using Equation 6, we obtain
\begin{equation}
    \Delta (\ln j) = \int_{(t_{\rm{obs}} - t_{\rm{quench}})}^{t_{\rm{obs}}} \int_{\mu_{min}}^1 \lambda(M_*,\mu,z)\ln F(\mu)~d\mu dt.
\end{equation}
Note that $M_*$ is also a function of time. Since we are considering the post-quenching evolution, we assume that the change in $M_*$ is entirely due to mergers; i.e., for each merger event,
$M_*^{\rm post}=(1+\mu)\,M_*^{\rm begin}$. The expected (i.e., averaged over $\mu$) logarithmic mass-growth rate is then
\begin{equation}
    \frac{d\ln M_*}{dt}(M_*,z) = \int_{\mu_{\rm{min}}}^{1} \lambda(M_*,\mu,z)\ln (1+\mu)~d\mu,
\end{equation}
and therefore
\begin{equation}
    \frac{d M_*}{dt}(M_*,z) = M_*\int_{\mu_{\rm{min}}}^{1} \lambda(M_*,\mu,z)\ln (1+\mu)~d\mu.
\end{equation}
The example $M_*$ evolution tracks are shown in the left panel of Figure~\ref{fig:J_model} for the two characteristic stellar masses considered in this work. Finally, the cumulative change in specific angular momentum after quenching can then be computed by combining Equations 19 and 21.

The right panel of Figure \ref{fig:J_model} shows the cumulative change in specific angular momentum since quenching, $\Delta(\ln j)$, as a function of $t_{\rm quench}$, with colors indicating stellar mass
at $z=0.8$. Two qualitative trends are evident.

First, at fixed stellar mass, the magnitude of the cumulative loss increases with $t_{\rm quench}$, as expected since the instantaneous loss rate of specific angular momentum is negative throughout the
post-quenching evolution (Section \ref{sec:J_loss_rate}). For short quenching ages, $\Delta(\ln j)$ is small (i.e., $j$ is only mildly reduced), whereas
for larger $t_{\rm quench}$ the reduction can become substantial, implying that even a modest instantaneous loss rate can
translate into an appreciable change in $j$ once integrated over several Gyr.

Second, the cumulative loss is strongly mass dependent: more massive systems exhibit systematically more negative
$\Delta(\ln j)$ at a given $t_{\rm quench}$. This follows directly from the left-panel result that $|d\langle \ln
j\rangle/dt|$ increases with $M_*$. In other words, for two galaxies that quenched at the same time, the higher-mass galaxy is expected to experience both a larger instantaneous loss rate and, consequently, a larger integrated loss by the time it is observed. This produces the separation between the colored curves, with the highest-mass tracks showing the largest cumulative decline.

We compare the predictions from the toy model to the observational results presented in this study. Using the
reconstructed SFHs from \prospector\ (Section \ref{sec:measure}), we estimate $t_{\rm quench}$ for each galaxy in our
sample by computing the lookback time at which the specific SFR (sSFR) drops below $10^{-10}\,\mathrm{yr}^{-1}$. We find
a median $t_{\rm quench}=3$ Gyr for our sample, consistent with measurements for LEGA-C galaxies \citep{Nersesian2025} and
similar to those for other samples of massive quiescent galaxies at comparable redshift \citep{Tacchella2022}. For the
youngest and oldest bins shown in Figure~\ref{fig:sigma_q}, the median quenching times are $t_{\rm quench}\approx 0.7$
and $5$ Gyr, respectively. 

Using the results from the right panel of Figure \ref{fig:J_model}, we compute the expected loss of specific angular
momentum across the three age bins in Figure \ref{fig:sigma_q} for two characteristic stellar masses,
$M_*=10^{11}\,M_\odot$ and $10^{11.7}\,M_\odot$, corresponding to the median stellar masses of the low- and high-mass bins,
respectively.

For the high-mass bin, the toy model predicts a substantial cumulative decline in $j$ with increasing $t_{\rm quench}$.
At $t_{\rm quench}=5$~Gyr, we obtain $\Delta(\ln j)\approx -1.5$, which corresponds to a fractional loss of
\begin{equation}
1-\frac{j(t_{\rm obs})}{j(t_{\rm obs}-5~{\rm Gyr})}
= 1-e^{-1.5} \approx 0.8,
\end{equation}
i.e., galaxies retain only $\sim 20\%$ of their specific angular momentum relative to the epoch immediately after
quenching. In contrast, for recently quenched systems with $t_{\rm quench}\simeq 0.7$~Gyr, we find
$\Delta(\ln j)\approx -0.1$, implying that such massive galaxies retain $\sim 90\%$ of their post-quenching $j$. The model
therefore predicts a strong trend in $j$ among the three age bins for the high-mass sample: older galaxies
should show systematically lower $j$ than younger galaxies, with the difference becoming dramatic once
$t_{\rm quench}\gtrsim {\rm few}$ Gyr.

For the low-mass bin, the predicted evolution is qualitatively similar but weaker in amplitude. At $t_{\rm quench}=5$ Gyr,
the model yields a smaller cumulative change, corresponding to $j(t_{\rm obs})/j(t_{\rm obs}-5~{\rm Gyr})\sim 0.5$,
i.e., a retention fraction of order $50\%$ (compared to 20\% predicted for the high-mass bin). For recently quenched systems ($t_{\rm quench}\simeq 0.7$~Gyr), the model implies a retention fraction close to $\sim 90\%$. 

Thus, while the toy model
predicts an age dependence of $j$ in both mass bins, the expected difference between age bins is significantly larger
for the high-mass galaxies. This is a direct consequence of the mass dependence of the merger-driven loss rate shown in
the left panel of Figure \ref{fig:J_model}: more massive descendants experience higher merger rates and therefore accrue larger cumulative losses in
$j$ over the same post-quenching interval.

Observationally, we
find that older quiescent galaxies exhibit lower $j$ than younger systems at fixed redshift  (Figure~\ref{fig:v_2_s}), consistent with the sign of
the model prediction. Moreover, the observed age dependence is more pronounced in the high-mass bin than in the low-mass
bin (Figure~\ref{fig:sigma_q}), qualitatively matching the toy-model expectation that merger-driven evolution should be most efficient for the most
massive galaxies.

\subsubsection{Caveats and limitations of the toy model} 

Our model for $\ln j$ is designed to capture qualitative trends with minimal bookkeeping. Therefore, it should not be interpreted as a precision predictor of angular-momentum evolution. The dominant caveat is that we effectively treat mergers as dry (i.e., gas-poor) and assume random relative orientations between the progenitors' internal spins and the merger orbital angular momentum. In what follows, we discuss the limitations of our toy model in detail.

First, because we focus on quiescent galaxies, treating mergers as predominantly dry is a reasonable zeroth-order approximation. In particular, observational reconstructions of star-formation histories indicate that star-formation rejuvenation episodes exist but contribute only a small fraction of the stellar mass budget of the quiescent population (e.g., a few per cent at $z\sim0.8$ in LEGA-C, \citealt{Chauke2019}), implying that dissipation-driven, post-quenching growth of high-$j$ stellar components is {\it not} the dominant channel for the ensemble angular-momentum evolution we study here. Nevertheless, even within a quiescent sample, residual gas, recently quenched satellites, or continuous (minor) gas accretion can partially preserve/rebuild rotation; in those cases our toy model, which neglects dissipation and in-situ regrowth, will tend to over-predict the decrease in $\ln j$. 

The other key idealization in the toy model is that merger orientations are random, so that the expected evolution reflects average vector cancellation (Equation 9). This assumption is motivated by the following three main points. First, simulations show that the $\ln j$ response of dry mergers depends strongly on alignment (misaligned dry mergers drive larger $j$ decreases, \citealt{Lagos2018}), making an angle-averaged treatment a natural first step. Second, observations in the nearby Universe found that massive early-type slow rotators exhibit broadly distributed (nearly flat) kinematic misalignment angles \citep{Davis2011,Ene2018}, consistent with external accretion/mergers arriving from a wide range of directions. Third, cosmological simulations show that mergers can induce substantial spin re-orientations \citep{Welker2014}, which helps erase memory of prior orientations and further supports a stochastic description at the population level. With all that being said, we caution, however, that accretion is not strictly isotropic: massive  systems can show statistical correlations with the cosmic web (e.g., a tendency for spins to be preferentially perpendicular to filaments, \citealt{GaneshaiahVeena2019}), implying that ``random'' is an approximation rather than a theorem. The direction of the resulting bias is therefore partly case-dependent: if post-quenching growth is preferentially aligned, cancellation is reduced and the toy model overestimates $j$-loss (under-predicts 
$\ln j$); if misaligned encounters are overrepresented, it can underestimate the true $\ln j$ loss.

Beyond these two assumptions, the toy model necessarily compresses the broad distribution of merger orbits (eccentricity, pericenter, impact parameter) and remnant structural response (non-homology, anisotropy, and finite relaxation time) into a single effective prescription, so object-by-object predictions can scatter widely even when the mean trend is captured. The toy model omits non-merger channels that can reshape the measured $j$ of quiescent systems without large rejuvenation, e.g., secular redistribution, stellar-halo/envelope growth, and environmental heating/stripping, and it is sensitive to operational choices in how $j$ is defined and measured (aperture, tracer, and projection), all of which can introduce systematic offsets in $\ln j$ when comparing to data or simulations.

We stress that our ability to confront the toy model with the observations presented in this work is fundamentally limited by the kinematic information available. Our measurements rely on unresolved  kinematics, so we cannot robustly recover spatially resolved rotation or obtain galaxy-by-galaxy estimates of $j$. Instead, we infer angular-momentum loss qualitatively from the empirical behavior in the \sigstar-$q$ plane (Section \ref{sec:sigma_q}), which is sensitive to the balance between ordered and random motions but does not uniquely map onto a specific $\Delta\ln j$. We have constructed a simple mapping from the \sigstar-$q$  diagram to an effective $V/\sigma$ (Section \ref{sec:model}), but this inference carries systematic uncertainties as mentioned in Section \ref{sec:model}. Ultimately, accurate estimates of $V/\sigma$ for individual galaxies will require IFU spectroscopy, which is not available for our sample and is unlikely to become available in the near term given the large sample size.

Finally, we note that the toy model for $\ln j$ is conceptually linked to changes in $V/\sigma$ via
\begin{equation}
d\ln j \;=\; d\ln (V/\sigma) \;+\; d\ln \sigma \;+\; d\ln R,
\end{equation}
where $R$ is the size of galaxies, but converting a modeled $d\ln j$ into a predicted $d\ln(v/\sigma)$ requires knowledge of the accompanying changes in $\sigma$ and $R$ (and, more generally, the structural and anisotropy evolution that sets the mapping between observables and intrinsic kinematics), which we do not directly constrain for our sample. Given these observational limitations on $V/\sigma$, together with the deliberately simplified nature of the $\ln j$ toy model, we therefore refrain from a fully quantitative comparison between model predictions and the data. Nevertheless, the key qualitative trends are consistent: the observed evolution in the \sigstar-$q$ plane matches the direction expected for angular-momentum loss, and the characteristic stellar-mass scale at which the model predicts strong $j$-loss is in broad agreement with the mass scale at which the data indicate a change in kinematic/shape behavior. Taken together, this concordance provides strong evidence that our observational results can be explained, if not uniquely, by dry merger-driven, post-quenching angular-momentum evolution in the quiescent population.

\section{Summary} 

We study the connection between quenching and dynamical transformation using unresolved stellar kinematics for 952 UVJ-selected quiescent galaxies with $M_\ast>10^{10.5}\,M_\odot$ at $0.6<z<1$ from LEGA-C. We combine integrated stellar velocity dispersions with projected axis ratios and Prospector-based stellar ages to examine how the empirical $\sigma'$--$q$ relation varies with time since quenching and with stellar mass.

A key empirical result is that the youngest quiescent populations show a clear dependence of $\sigma'$ on $q$, consistent with a non-negligible contribution from ordered rotation to the integrated line broadening ((Section \ref{sec:sigma_q}). This indicates that galaxies can enter the quiescent population while still retaining substantial rotational support; i.e., quenching does not necessarily coincide with an immediate transition to a fully dispersion-supported state.

We then find that the subsequent evolution is strongly mass dependent (Section \ref{sec:sigma_q}). At lower stellar masses ($<10^{11.3}M_\odot$), the $\sigma'$--$q$ behavior remains broadly similar across stellar-age bins, consistent with continued rotational support several Gyr after quenching. In contrast, at the highest masses ($>10^{11.3}M_\odot$), the $\sigma'$--$q$ relation becomes progressively flatter with increasing stellar age, consistent with a gradual reduction of net ordered rotation within the massive quiescent population.

To connect these trends to an intuitive measure of dynamical support, we introduce a simple toy model that maps the observed $\sigma'$--$q$ relation onto an effective $V/\sigma$. The model reproduces the median trends and suggests a stronger post-quenching decline of rotational support at higher mass. We stress, however, that the \textbf{absolute} $V/\sigma$ values are model dependent: assumptions such as the conversion factor $\gamma$, intrinsic thickness $q_z$, and related parameters can shift the inferred normalization (Section~\ref{sec:model}).

To interpret the implied angular-momentum evolution, we additionally develop a simple stochastic framework in which the cumulative effect of incoherent accretion/merging events produces approximately additive changes in $\ln j$ (Section \ref{sec:loss_J}). Despite its simplicity, this toy model captures the key qualitative physics of repeated, misaligned perturbations that reduce net ordered rotation, and it yields trends that are qualitatively consistent with our data, namely, a stronger cumulative angular-momentum loss at higher stellar mass and a gradual decline of rotational support within the massive quiescent population over time.

Our results are consistent with a picture in which angular-momentum loss can proceed continuously within the quiescent population after quenching, rather than occurring instantaneously at the time quenching. This interpretation is also supported by the small but growing set of spatially resolved stellar-kinematic measurements of quiescent galaxies at $z\sim2$, particularly strongly lensed systems, which show that at least some very massive, recently quenched galaxies can retain substantial rotational support (high $V/\sigma$) at early times. 

Finally, our $z\sim0.8$ analysis, based on spatially unresolved spectroscopy, provide earlier evidence that the mass-dependent kinematic transformation associated with the fast/slow rotator dichotomy is already operating by $z\sim0.8$. Whether the dichotomy is already largely established by $z\sim0.8$, or still emerging at that time, remains uncertain and will require spatially resolved stellar kinematics at intermediate redshift, ideally coupled with comparisons to cosmological simulations, to pin down the dominant physical pathways.

\begin{acknowledgments}
This work was completed in part with resources provided by the Green High Performance Computing Cluster (GHPCC) of the University of Massachusetts Amherst.
\end{acknowledgments}

\software{Prospector \citep{Johnson2021}, FSPS \citep{Conroy2009,Conroy2010}, MIST \citep{Choi2016,Dotter2016}, MILES \citep{Falcon-Barroso2011}, GALFIT \citep{Peng2010}} 

\appendix
\vspace{-27pt}
\section{Testing the robustness of the stellar-age inference} \label{app:age}

Here we present and discuss in detail about our tests on the stellar-age measures from our \prospector fitting. 

To begin, we compare the best-fit SED models predicted by \prospector with the observed LEGA-C spectra. Specifically, we compare the median observed and predicted spectra of each one of the subgroups presented in the main text (Section \ref{sec:result}). We remind that our \prospector fitting only used photometric data ($\approx$ 40 bands), i.e. the spectra were not used in the SED modeling. This comparison thus allows us to have a direct and broad view on the quality of our SED fitting. 

\begin{figure*}
    \centering
    \includegraphics[width=0.97\textwidth]{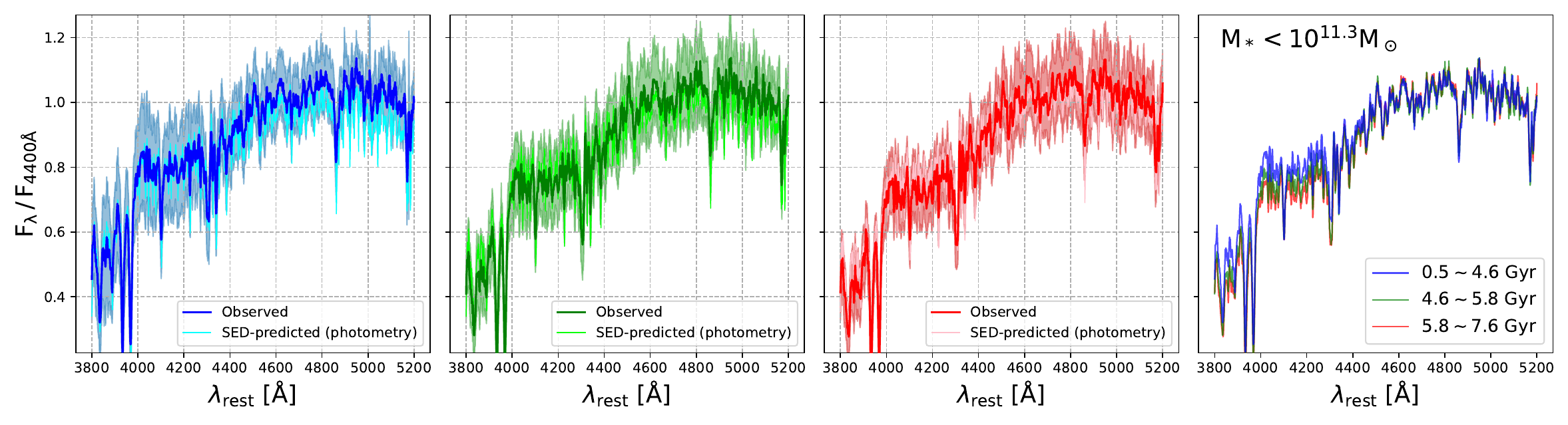}
    \includegraphics[width=0.97\textwidth]{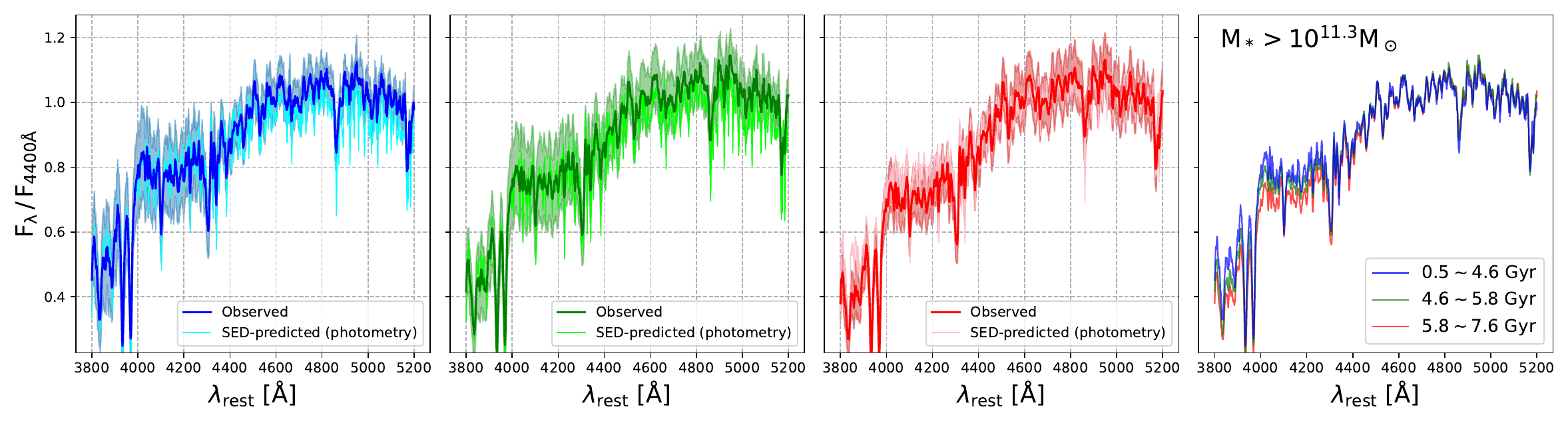}
    \caption{Comparison between the median observed (LEGA-C) and predicted (\prospector fitting with photometry only) spectra of each one of the subgroups discussed in the main text. The sample division here is the same as used in Figure \ref{fig:sigma_q}. In each panel, we also show the 1-$\sigma$ range (standard deviation) of the observed spectrum as shaded region. In the right-most panel, the median observed spectra of all stellar age bins are shown together.  }
    \label{fig:obs_vs_pre}
\end{figure*}

As Figure \ref{fig:obs_vs_pre} shows, the median spectra of all subgroups dim at the rest-frame UV wavelengths, and show strong stellar absorption features without any strong emission lines over rest-frame $3800-5200$ \AA, which demonstrates, once again, the effectiveness of the UVJ technique in identifying quiescent galaxies.
The the best-fit spectra from SED modeling are in excellent agreement with the observed ones, within the uncertainties, for all subgroups. Moreover, as the right-most panel of Figure \ref{fig:obs_vs_pre} shows, galaxies with older stellar ages -- inferred from \prospector -- also have redder observed spectra, even though the spectral information was not included during the SED fitting procedure. This  agreement suggests that the stellar-age inference from our SED fitting procedure is robust. We stress, however, that the conclusion above {\it does not at all} imply that the spectral information is not needed for SED fitting. In fact, robust measurements of metallicity and elemental abundance are only possible with spectra. What we really mean is that the stellar age of high-z quiescent galaxies can be inferred robustly when densely-sampled, panchromatic photometry is available. Similar conclusions were also reached by \citet{paperi} using synthetic galaxies from cosmological simulations. 

Despite the numerous advantages of fitting photometric and spectral data simultaneously \citep[e.g.,][]{Tacchella2022}, here we want to highlight one potential, serious systematics: the non-trivial aperture matching that is required when combining photometry and spectroscopy. To tackle this, most studies simply rescale (or perform the fit with the rescaling factor as a free parameter) the observed spectra to the same flux level of photometry. The big assumption behind such a procedure is that there is no strong color/stellar-population variation between the photometric aperture and the spectral slit. This assumption can be problematic given that color gradients have been clearly observed in high-z massive quiescent galaxies \citep[e.g.][]{Suess2020,Ji2024}. Ideally, in order to simultaneously fitting photometry and spectroscopy, one needs forward modeling the instrumental effects to properly account for e.g. the mismatch of apertures, which however is beyond the scope of this work. The fact that the predicted spectra using photometry alone are in very good agreement with the observed ones provides confidence that the measures of stellar age of the sample galaxies are robust.

\begin{figure*}
    \centering
    \includegraphics[width=0.47\textwidth]{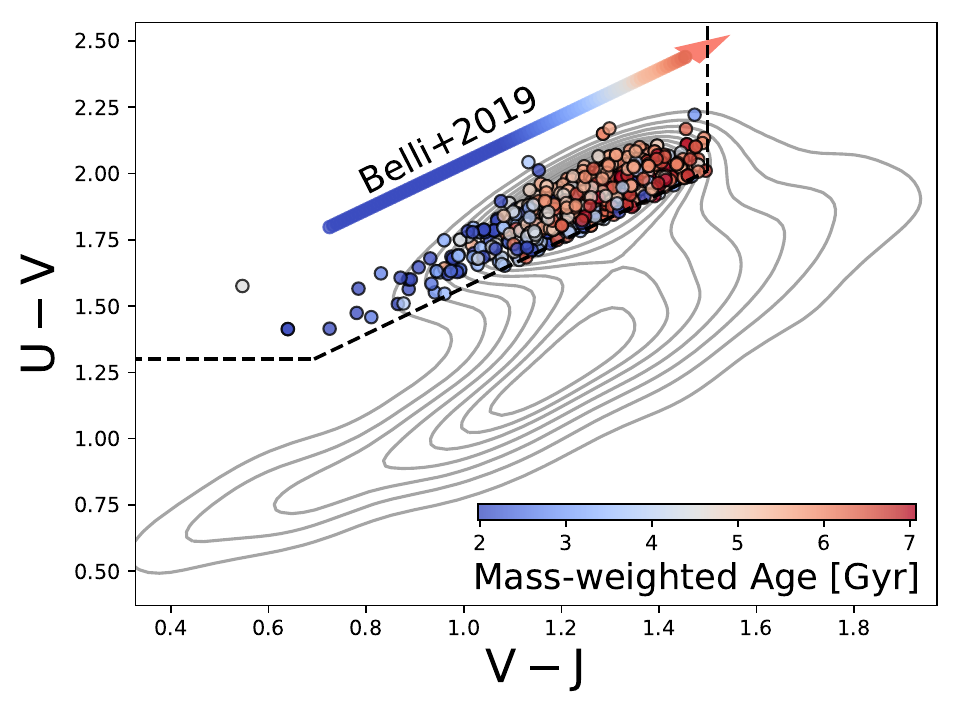}
    \includegraphics[width=0.47\textwidth]{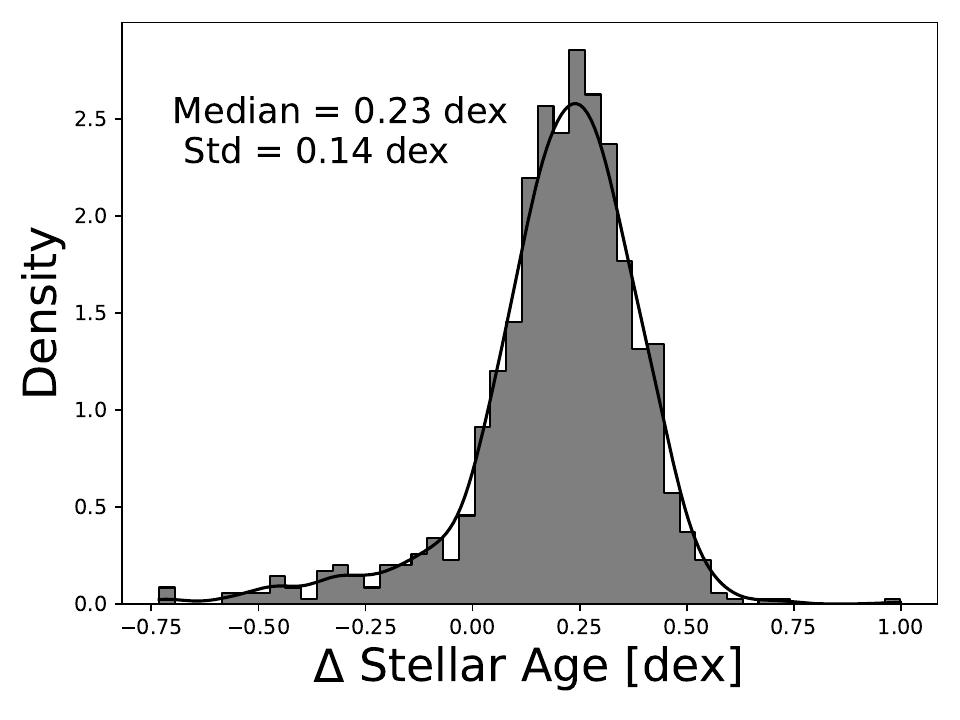}
    \caption{{\bf Left:} UVJ diagram. The quiescent galaxies of this study are shown as individual circles color coded according to stellar ages from our \prospector fitting. The background contours show the distribution of all LEGA-C galaxies with $M_*>10^{10.5}M_\sun$. The black dashed lines mark the UVJ selection criteria of \citet{Muzzin2013}. The vector shows the best-fit relation between the stellar age and UVJ colors from \citet{Belli2019}. Great qualitative agreement -- galaxies with bluer UVJ colors are younger -- is seen between the two stellar age inferences. {\bf Right:} Distribution of the difference in stellar age between our \prospector measurements and those inferred using the best-fit relation of \citet{Belli2019}. Our measurements on average return a $0.23\pm0.14$ dex older stellar age.}
    \label{fig:age_compare}
\end{figure*}

We continue our tests on the stellar-age measures by comparing the age-color relationship from our \prospector fitting with earlier studies. \citet{Whitaker2013} stacked the 3D-HST grism spectra of massive quiescent galaxies at $z\sim2$, and fit the stacked spectral features with a solar-metallicity, single stellar population model. They found that quiescent galaxies having bluer rest-frame $(U-V)$ and $(V-J)$ colors are younger than those having the redder colors. We made a similar comparison and found excellent qualitative agreement, as we show in Figure \ref{fig:age_compare}.  

Similarly, \citet{Belli2019} derived and calibrated the relationship between stellar age and rest-frame UVJ colors using a sample of 24 quiescent galaxies at $1.5<z<2.5$ with deep rest-optical spectroscopy.  The stellar ages of their sample galaxies were estimated by modeling spectra and photometry combined. To assess the consistency between their results and ours, we compare the relationship between stellar age and UVJ colors of our measurements with that of \citet{Belli2019}. To do so, we first derive the quiescent sequence of our sample by conducting a linear regression between the $(U-V)$ and $(V-J)$ colors. With the best-fit linear relationship between the two colors in hand, we use Equation 3 of \citet{Belli2019}\footnote{$\log(\rm{Age/yr})=7.03 + 0.84\, (V - J) + 0.74\, (U - V)$} to derive the expected change of stellar age along our quiescent sequence, which is shown as the vector in Figure \ref{fig:age_compare}. We run a Pearson correlation test between the stellar ages from \prospector and the stellar ages inferred using the best-fit relation of \citet{Belli2019}. We found a strong correlation, with Pearson coefficient $r=0.52$), which demonstrates good qualitative agreement between our stellar-age measures and those of \citet{Belli2019}. Note, however, that the above relationship \citep{Belli2019} between age and rest-frame colors was derived at $z\sim 1.7$. As the galaxies age and their colors evolve, we expect the relationship to change as well. While the study of the color evolution of quiescent galaxies is beyond the scope of this work, we note that, as the right panel of Figure \ref{fig:age_compare} shows, a systematic age difference is observed between our measures with \prospector at $z\sim0.8$ and those of \citet{Belli2019} at $z\sim1.7$ such that the $z\sim0.8$ galaxies are older than those at $z\sim1.7$ by 0.23 dex, or 2.6 Gyr which is about the time interval between the two redshifts, $t_{\rm{H}}^{z=0.8}-t_{\rm{H}}^{z=1.7}\approx2.9$ Gyr. 

In terms of interpretation, this systematic offset primarily affects the \emph{absolute} age scale, rather than the \emph{relative} age ranking of galaxies within our $0.6<z<1$ sample. A roughly uniform shift in the age zero-point would move the inferred ages of all bins by a similar amount, but would not change the ordering of ``younger'' versus ``older'' quiescent populations at fixed redshift. Therefore, our main conclusions, in particular the mass-dependent evolution of kinematic support inferred from the changing $\sigma'_\star$--$q$ relation with stellar age, are robust to such an offset. The primary impact is on statements that translate the age axis into an absolute timescale for angular-momentum loss; these should be regarded as approximate (at the Gyr level) given cross-method systematics.

In summary, by comparing (1) the predicted spectra from \prospector with the observed ones from LEGA-C and (2) the stellar ages from \prospector and the ones inferred using the age-color relationship reported by previous studies, we showed very good agreement among different stellar-age measures. We stress that, because the conclusions of this study only depend on differential stellar-age measures (younger vs. older), rather than absolute stellar-age determinations, we conclude that the results and key conclusions presented in this study are robust.  

\bibliography{ji_2022_legac}
\end{document}